\documentclass[aps, prl, twocolumn, nofootinbib, floatfix, superscriptaddress]{revtex4-1}

\usepackage{graphicx}% Include figure files
\usepackage{dcolumn}% Align table columns on decimal point
\usepackage{bm}% bold math
\usepackage{subfigure}
\usepackage{epsfig}
\usepackage{amsmath}
\usepackage{amsfonts}
\usepackage{graphicx}
\usepackage{amssymb}
\usepackage{nccmath}
\usepackage{amssymb,amsmath,amsfonts}
\usepackage{xfrac}
\usepackage{color}
\usepackage{tikz}
\usepackage[T1]{fontenc}
\usepackage[utf8]{inputenc}
\usepackage{ulem}

\usepackage{tcolorbox}
\usepackage{hyperref}
\hypersetup{colorlinks=true, citecolor=red, linkcolor=blue, urlcolor=blue}

\definecolor{rossos}{cmyk}{0,1,1,0.55}
\definecolor{bluscuro}{rgb}{0.15, 0.2, .85}
\definecolor{bluchiaro}{cmyk}{1,.3,0.,0.1}

\newcommand{\OmegaGW}{\Omega_{\mathrm{GW}}}

\let\oldsqrt\sqrt
\def\sqrt{\mathpalette\DHLhksqrt}
\def\DHLhksqrt#1#2{%
\setbox0=\hbox{$#1\oldsqrt{#2\,}$}\dimen0=\ht0
\advance\dimen0-0.2\ht0
\setbox2=\hbox{\vrule height\ht0 depth -\dimen0}%
{\box0\lower0.4pt\box2}}

\newcommand{\sss}[1]{{\scriptscriptstyle{#1}}}
\newcommand{\boldmathsymbol}[1]{{\ensuremath{\boldsymbol{#1}}}}

\newcommand{\uPl}{\mathrm{Pl}}

\newcommand{\usssPl}{\sss{\uPl}}

\newcommand{\calH}{\mathcal{H}}

\newcommand{\Mp}{M_\usssPl}

\newcommand{\beq}{\begin{equation}}
\newcommand{\eeq}{\end{equation}}
\newcommand{\bea}{\begin{equation}\begin{aligned}}
\newcommand{\eea}{\end{aligned}\end{equation}}

\newlength{\wsingfig}
\newlength{\wdblefig}
\newlength{\wquadfig}
\newlength{\wtriplefig}
\newcommand{\Eq}[1]{Eq.~(\ref{#1})}

\newcommand{\be}{\begin{equation}}

\begin{document}

 \title{Induced gravitational waves from flipped SU(5) superstring theory at $\mathrm{nHz}$}

 \author{Spyros Basilakos}
% \email{svasil@Academyofathens.gr}
\affiliation{National Observatory of Athens, Lofos Nymfon, 11852 Athens, 
Greece}
\affiliation{Academy of Athens, Research Center for Astronomy and Applied 
Mathematics, Soranou Efesiou 4, 11527, Athens, Greece}
\affiliation{School of Sciences, European University Cyprus, Diogenes 
Street, Engomi, 1516 Nicosia, Cyprus}

\author{Dimitri~V.~Nanopoulos}
%\email{dimitri@physics.tamu.edu}
\affiliation{George P. and Cynthia W. Mitchell Institute for Fundamental
 Physics and Astronomy, Texas A\&M University, College Station, Texas 77843, 
USA}
 \affiliation{Astroparticle Physics Group, Houston Advanced Research Center 
(HARC),
  Mitchell Campus, Woodlands, Texas 77381, USA }
\affiliation{
  Academy of Athens, Division of Natural Sciences,
Athens 10679, Greece }

\author{Theodoros Papanikolaou}
%\email{t.papanikolaou@ssmeridionale.it}

\affiliation{Scuola Superiore Meridionale, Largo S. Marcellino 10, I-80138, Napoli, Italy}
\affiliation{National Observatory of Athens, Lofos Nymfon, 11852 Athens, 
Greece}

\author{Emmanuel N. Saridakis}
%\email{msaridak@noa.gr}
 \affiliation{National Observatory of Athens, Lofos Nymfon, 11852 Athens, 
Greece}
\affiliation{CAS Key Laboratory for Researches in Galaxies and Cosmology, 
Department of Astronomy, University of Science and Technology of China, Hefei, 
Anhui 230026, P.R. China}

\author{Charalampos Tzerefos}
%\email{chtzeref@phys.uoa.gr}
 \affiliation{National Observatory of Athens, Lofos Nymfon, 11852 Athens, 
Greece}
 \affiliation{Department of Physics, National \& Kapodistrian University of 
Athens, Zografou Campus GR 157 73, Athens, Greece}

%\pacs{98.80.-k, 95.36.+x }

\begin{abstract}
The no-scale flipped SU(5) superstring framework constitutes a very promising paradigm for physics below the Planck scale providing us with a very rich cosmological phenomenology in accordance with observations. In particular, it can accommodate Starobinsky-like inflation, followed by a reheating phase, which is driven by a light "flaton" field, and during which the GUT phase transition occurs. In this Letter, we extract for the first time a gravitational-wave (GW) signal which naturally arises in the context of the flipped SU(5) cosmological phenomenology and is related to the existence of an early matter era (eMD) driven by the flaton field. Specifically, we study GWs non-linearly induced by inflationary perturbations and which are abundantly produced during a sudden transition from the flaton-driven eMD era to the late-time radiation-dominated era. Remarkably, we find a GW signal with a characteristic peak frequency $f_\mathrm{GW,peak}$ depending only on the string slope $\alpha'$ and reading as $f_\mathrm{GW,peak} \propto 10^{-9} \left(\frac{\alpha'}{\alpha'_*}\right)^4 \mathrm{Hz}$, where $\alpha'_*$ is the fiducial string slope being related directly to the reduced Planck scale $\Mp$ as $\alpha'_* = 8/\Mp^2$. Interestingly enough, $f_\mathrm{GW,peak}$ lies within the $\mathrm{nHz}$ frequency range; hence rendering this primordial GW signal potentially detectable by SKA, NANOGrav and PTA probes at their very low frequency region of their detection bands.
\end{abstract}

\maketitle

%\section{Introduction}

%%%%%%%%%%%%%%%%%%%%%%%% Section 1:  Introduction %%%%%%%%%%%%%%%%%%%%%%%%%%%%%%%%%%%%%%%%%%%%%%%%%%%%%%%%%%%%%%%%
{\bf Introduction } -- 
Superstring theory constitutes one of the most promising paradigms describing the underlying fundamental quantum "theory of everything" including gravity as well as the standard model (SM). If string theory describes nature unifying all fundamental interactions, it should be able to explain all physical phenomena at all scales. String phenomenology has made several steps towards this goal by string constructions of particle physics models and continue then with concrete models of cosmology \cite{Antoniadis:1989zy,Rizos:1990xn,Antoniadis:1991fc,Cvetic:1998gv,Cleaver:1997nj,Cicoli:2023opf}. Though, very little has been done in explicit constructions of both at the same time. 

In this Letter, we focus on the flipped SU(5) superstring framework which can give us a simultaneous description of particle phenomenology and cosmology~\cite{Antoniadis:2020txn,Antoniadis:2021rfm}. In particular, it can naturally give rise to Starobinsky-like inflationary setups~\cite{Ellis:2013xoa,Ellis:2013nxa}, favored by CMB observations, and explain the neutrino mass generation through an inflaton-induced see-saw mechanism~\cite{Ellis:2013nka,Ellis:2017jcp}, the baryon asymmetry in the Universe as well as the neutralino dark matter \cite{Goldberg:1983nd,Ellis:1983ew}. 

Interestingly enough, during inflation, the vacuum remains in the unbroken
GUT phase, and GUT symmetry breaking occurs later when a field with a flat direction (the flaton)
acquires a vacuum expectation value. At the end, reheating is driven by the incoherent part of the flaton operating in the so-called strong regime~\cite{Ellis:2017jcp,Ellis:2018moe,Ellis:2019opr}. As the Universe cools down, the flaton becomes non-relativistic and eventually triggers an early matter-dominated era (eMD) \cite{CAMPBELL1987355, Ellis:2018moe,Ellis:2019opr}.

During this eMD era~\cite{1981SvA....25..406P,1982AZh....59..639P}, the inflationary adiabatic perturbations can be significantly enhanced inducing at second order in cosmological perturbation theory an abundant production of gravitational waves~\cite{Inomata:2019ivs, Inomata:2019zqy, Inomata:2020lmk, Papanikolaou:2020qtd, Domenech:2020ssp, Papanikolaou:2022chm}. These scalar-induced GWs (SIGWs), behaving as radiation, can contribute as well to the effective number of extra neutrino species $\Delta N_\mathrm{eff}$ potentially relaxing in a natural way the Hubble tension. Interestingly enough, as it was shown in~\cite{Bhaumik:2022zdd,Papanikolaou:2023oxq}, enhanced cosmological perturbations can collapse and form primordial black hole (PBHs) inducing as well GWs that can increase $\Delta N_\mathrm{eff}$, leading in this way to the alleviation of the Hubble tension.

It is these GWs that we study within this work. Remarkably, we find a peak GW frequency $f_\mathrm{GW,peak}$ which depends only on the string slope $\alpha'$ and reading as $f_\mathrm{GW,peak} \propto 10^{-9} \left(\frac{\alpha'}{\alpha'_*}\right)^4 \mathrm{Hz}$, where $\alpha'_*$ is the fiducial string slope being related to the reduced Planck scale $\Mp$ as $\alpha'_* = 8/\Mp^2$. Notably, $f_\mathrm{GW,peak}$ lies within the $\mathrm{nHz}$ frequency range; hence rendering this primordial GW signal detectable by SKA, NANOGrav and PTA probes.

%%%%%%%%%%%%%%%%%%%%%%%% Section 2:  Flipped SU(5) Superstring paradigm %%%%%%%%%%%%%%%%%%%%%%%%%%%%%%%%%%%%%%%%%%%%%%%%%%%%%%%%%%%%%%%%
{\bf The flipped SU(5) superstring paradigm}-- The string-derived flipped SU(5) model, initially introduced in~\cite{Antoniadis:1989zy}, constitutes a step towards a unified particle and cosmological phenomenology~\cite{Antoniadis:2020txn,Antoniadis:2021rfm} being built within the framework of the free fermionic formulation (FFF) of four-dimensional (4d) heterotic superstrings \cite{ANTONIADIS198787, ANTONIADIS1988586, Kalara:1990sq}. The FFF has the advantage of calculability in perturbation theory around a vacuum where all moduli, except the dilaton, are fixed at the fermionic point by a set of gauge and discrete symmetries, such as flipped $\mathrm{SU(5)} \times \mathrm{U(1)}$ \cite{BARR1982219,DERENDINGER1984170,ANTONIADIS1987231}.

The choice of the sub-Planckian gauge group, i.e. flipped $\mathrm{SU(5)} \times \mathrm{U(1)}$, is not accidental. In weakly-coupled heterotic string compactifications the matter representations are limited in size, e.g. to $\bar{5}$ and $\bar{10}$ representations of  $\mathrm{SU(5)}$ \cite{Ellis:1990sr}. This consideration motivates our choice for a $\mathrm{GUT}$ gauge group, namely the flipped $\mathrm{SU(5)} \times \mathrm{U(1)}$ \cite{BARR1982219,DERENDINGER1984170,ANTONIADIS1987231} one, pretty uniquely as it can be broken down to the Standard Model (SM) $\mathrm{SU(3)}_C \times \mathrm{SU(2)}_L \times \mathrm{U(1)}_Y$ group by a combination of $\bar{10}$ and $10$ antisymmetric Higgs representations, whereas $\mathrm{SU(5)}$ and larger $\mathrm{GUT}$ groups require adjoint or larger Higgs representations. The resemblance of the representation content of $\mathrm{SU(5)} \times \mathrm{U(1)}$ and that of the SM $\mathrm{SU(3)}_C \times \mathrm{SU(2)}_L \times \mathrm{U(1)}_Y$ is rather striking. Interestingly enough, flipped $\mathrm{SU(5)} \times \mathrm{U(1)}$ offers a promising framework for supersymmetric grand unification that provides resolutions of several phenomenological issues in particle physics. For example, in addition to accommodating, naturally, light neutrino masses, it provides a minimal mechanism for splitting the masses of the triplet and doublet components of the fiveplets of Higgs fields and in such a way that avoids automatically  the dimension-5 operator that leads to catastrophically fast proton decay, endemic in supersymmetric $\mathrm{GUTs}$, like $\mathrm{SUSY} \, \mathrm{SU(5)}$ \cite{ANTONIADIS1987231}. 

Recently, the particle physics and cosmology of the string derived flipped $\mathrm{SU(5)} \times \mathrm{U(1)}$ have been studied in great detail \cite{Antoniadis:2020txn, Antoniadis:2021rfm}. Specifically, a set of vacuum expectation values (vevs) $\langle \phi_i \rangle $ are triggered by the breaking of an anomalous $\mathrm{U(1)}_A$  gauge symmetry \cite{Dine:1987xk}, $\langle \phi_i \rangle \sim \xi$ of roughly an order of magnitude below the string scale, which are induced in order to satisfy the  $F-$ and $D-$ flatness superpotential conditions up to 6th order ~\cite{Antoniadis:2021rfm}.
In particular, all extra color triplets become superheavy, guarantying observable proton decay stability while the Higgs doublet mass matrix has a massless pair eigenstate with realistic hierarchical Yukawa couplings to quarks and leptons \cite{Antoniadis:2021rfm}. Actually, more than thirty years ago, this string derived flipped $\mathrm{SU(5)} \times \mathrm{U(1)}$ model \cite{Antoniadis:1989zy} predicted that the mass of the top quark would be around $\sim 170-180 \mathrm{GeV}$ (see fig.1 of \cite{Antoniadis:1989zy}) as was observed at Fermilab in 1995. The stringy derived masses of quarks and leptons and their relations (at the string scale) are summarized below \cite{Antoniadis:2021rfm} e.g. 
\begin{align}
 & m_c \sim \xi^2 m_t, \,  m_s \sim \xi^2 m_b, \nonumber \\&  \,  m_u \sim \xi^5 m_t, m_d \sim \xi^3 m_b, \, m_e \sim \xi^4 m_t \, \, 
\end{align}
with $\xi$ dynamically calculated to be  $\xi = M_s/2\pi$  with  $M_s$ the string scale $M_s \equiv 1/\sqrt{2a'}$, with $a'$ the string slope, and thus, in \textit{string units}, $ \xi \approx \mathcal{O}(1/10)$. 

%%%%%%%%%%%%%%%%%%%%%%%% Section 3:  The flipped SU(5) superstring cosmology %%%%%%%%%%%%%%%%%%%%%%%%%%%%%%%%%%%%%%%%%%%%%%%%%%%%%%%%%%%%%%%%

{\bf The flipped SU(5) superstring cosmology}-- It has been shown~\cite{Antoniadis:2020txn} that the string derived no-scale flipped $\mathrm{SU(5)} \times \mathrm{U(1)}$  can accommodate also a successful cosmology, based on the no-scale supergravity~\cite{CREMMER198361,Ellis:1983sf,ELLIS1984406,ELLIS1984373,LAHANAS19871} derived from string theory~\cite{Antoniadis:2020txn,ANTONIADIS198796,FERRARA1987368,FERRARA1987358} as well as an appropriate induced superpotential suppressed  by five powers of the string scale~\cite{Antoniadis:2020txn}. It utilises two gauge singlet chiral superfields present in the low energy spectrum, namely the inflaton field $y$, identified as the superpartner of a state mixed with right-handed neutrinos, and the goldstino field $z$ with a superpotential of the form $W_\mathrm{I} = M_\mathrm{I}z (y- \lambda y^2)$, where $\lambda$ is a dimensionless $\mathcal{O}(1)$ parameter. $M_\mathrm{I} \sim 10^{13} \mathrm{GeV}$ is the mass scale of inflation, being generated at the 5th order by the breaking of an anomalous $\mathrm{U(1)}_\mathrm{A}$ gauge symmetry of the heterotic string chiral vacuum \cite{Antoniadis:2020txn}. \footnote{Note that $M_I \simeq C_6 a^5_s M_s \sim 10^{13} \mathrm{GeV}$ and $\lambda \simeq g_s \frac{M_\mathrm{Pl}}{M_s} \sim \mathcal{O}(1)$, with $C_6$ the numerical value of the correlator associated to the $N=6$ coupling and $a_s \equiv g_s/2\pi$, with $g_s$ the string coupling constant.} The resulting scalar potential leads to Starobinsky-type inflation with a scalar spectral index $ n_\mathrm{s} \simeq 0.965$ and a tensor-to-scalar ratio $ r= 2.4 \times 10^{-3}$ \cite{Antoniadis:2020txn} in excellent agreement with Planck~\cite{Planck:2018vyg}.

Furthermore, it should be stressed that during inflation and subsequent reheating, the $\mathrm{GUT}$ symmetry $\mathrm{SU(5)} \times \mathrm{U(1)}$ is left unbroken \cite{Ellis:2017jcp, Ellis:2018moe, Ellis:2019opr, ELLIS2019134864}. The $\mathrm{SU(5)} \times \mathrm{U(1)}$ symmetry is broken by a $10$ and $\bar{10}$ Higgs fields, $H$ and $\bar{H}$ respectively, along the $F-$ and $D-$ flat directions $\langle \tilde{\nu}^{c}_{H} \rangle =  \langle \tilde{\nu}^{c}_{\tilde{H}} \rangle  \neq 0$, where $\nu^{c}_{H}$ and $\nu^{c}_{\tilde{H}}$ are the SM singlets in the flipped $10$ and $\bar{10}$ representations respectively. These vevs, which can be naturally large thanks to the $F-$ and $D-$ flat directions, are induced by the soft supersymmetry breaking masses. The resulting symmetry breaking pattern is $\mathrm{SU(5)} \times \mathrm{U(1)} \rightarrow \mathrm{SU(3)}_C \times \mathrm{SU(2)}_L \times \mathrm{U(1)}_Y$.~\footnote{Notice that this symmetry breaking pattern is \textit{unique}, contrary to the case of SUSY $\mathrm{GUTs}$, which have degenerate vacua, e.g. $\mathrm{SU(5)}$ may be broken into other gauge groups, as $\mathrm{SU(4)} \times \mathrm{U(1)}$ \cite{Nanopoulos:1982zp}. We also mention that flipped  $\mathrm{SU(5)} \times \mathrm{U(1)}$ is free from any monopole problem, even if it is broken after inflation,  since the $\mathrm{SU(5)} \times \mathrm{U(1)}$ group is not simple \cite{DERENDINGER1984170}.} After $H$ and $\bar{H}$ develop vevs, a SM singlet, a linear combination of $\nu^{c}_{H}$ and $\nu^{c}_{\tilde{H}}$, appears as a physical state, massless in the supersymmetric limit, due to the presence of the $F-$ and $D-$flat direction in the potential, and has a mass of the order of the soft supersymmetric- breaking mass scale. We denote this combination by $\Phi$ and refer to it as the \textit{flaton} \cite{Ellis:2017jcp,Ellis:2018moe,Ellis:2019opr}. At low energies the soft supersymmetric breaking scale, i.e. the flaton mass-squared, $m_\mathrm{\Phi}^2$, is driven negative by renormalisation group equation effects (RGE) due to relevant Yukawa couplings \cite{ANTONIADIS1987231}. This negative mass-squared term destabilizes the origin of the flat direction, and thus the flaton field develops a vev, breaking the  $\mathrm{SU(5)} \times \mathrm{U(1)}$ $\mathrm{GUT}$ symmetry into the SM group.

Notice that if the height of the potential barrier between the symmetric ($\mathrm{SU(5)} \times \mathrm{U(1)}$) $\mathrm{GUT}$ and the Higgs ($\mathrm{SU(3)}_C \times \mathrm{SU(2)}_L \times \mathrm{U(1)}_Y$) phase of the theory is smaller than the thermal kinetic energy of $\Phi$, the incoherent part of $\Phi$ will drive the transition \cite{CAMPBELL1987355,Ellis:2018moe}, destroying any coherent contribution and displacing $\Phi$ to the low-energy vacuum within a Hubble time \cite{CAMPBELL1987355,Ellis:2018moe}. We call this scenario strong reheating \cite{Ellis:2017jcp,Ellis:2018moe,Ellis:2019opr}, during which a substantial amount of entropy release of $\mathcal{O}(10^4)$ reduces the cosmological baryon asymmetry and the density of the neutralino dark matter so as to be compatible with Planck \cite{Planck:2018vyg}. In particular, as the Universe cools down, the energy density of $\Phi$ will simply redshift as radiation, until the temperature of the Universe decreases at such level that the interactions keeping the flaton in thermal equilibrium cease to be efficient. At even later times, when $ m_\Phi > T_\Phi$, the  flaton becomes non-relativistic and eventually dominates the energy budget of the universe, leading to an early matter dominated era (eMD)~\cite{Ellis:2018moe,Ellis:2019opr}. 

This transient eMD era lasts until the end of the flaton decay, which proceeds fast~\cite{Ellis:2018moe, Ellis:2019opr, CAMPBELL1987355}. Consequently, the transition from the eMD era to the subsequent late radiation dominated era (lRD) is quite sudden. This means that the gravitational waves induced by inflationary adiabatic perturbations around the time of the flaton decay will become significantly enhanced due to the fast oscillations of the scalar metric perturbations well within the Hubble horizon \cite{Inomata:2019ivs}. It is the spectrum of these GWs that we are going to investigate in what follows.   

%%%%%%%%%%%%%%%%%%%%%%%% Section 4: Scalar induced gravitational waves %%%%%%%%%%%%%%%%%%%%%%%%%%%%%%%%%%%%%%%%%%%%%%%%%%%%%%%%%%%%%%%%
{\bf Scalar induced gravitational waves}--
Having described above the flipped $\mathrm{SU(5)}$ cosmological phenomenology, let us now review the basic formalism of the gravitational waves induced by second order gravitational interactions~\cite{Matarrese:1992rp,Matarrese:1993zf,Matarrese:1997ay,Mollerach:2003nq} [See~\cite{Domenech:2021ztg} for a review] and which in our setup are abundantly produced due to the existence of an eMD era driven by the flaton field. We choose to work within the so-called Newtonian gauge\footnote{Here, one should refer to the issue of gauge dependence of GWs emitted at second order in cosmological perturbation theory firstly studied in~\cite{Hwang:2017oxa}. As it was shown in~\cite{Tomikawa:2019tvi,DeLuca:2019ufz,Inomata:2019yww,Domenech:2020xin}, the gauge invariance of the second-order GWs is generically retained when the emission is followed by a phase where the GW source is not active anymore. In our case, although the GW emission takes place during an eMD era driven by the flaton field, during which the scalar and the tensor modes are coupled to each other, it is followed by the late RD era, during which the scalar perturbations decay very quickly and decouple from the tensor perturbations~\cite{Inomata:2019yww,Domenech:2020xin}. Thus, the GW signal computed here in the Newtonian gauge during the late RD era is gauge-independent.}, in which the perturbed metric is written as
\beq
\mathrm{d}s^2 = a^2(\eta)\left\lbrace-(1+2\Psi)\mathrm{d}\eta^2  + 
\left[(1-2\Psi)\delta_{ij} + 
\frac{h_{ij}}{2}\right]\mathrm{d}x^i\mathrm{d}x^j\right\rbrace,
\eeq
where $\Psi$ is the first order Bardeen gravitational potential and $h_{ij}$ denotes the 
second order tensor perturbation. Then, after a Fourier transformation of the tensor perturbation and minimisation of the tensor part of the cubic order gravitational action, one straightforwardly gets the equation of 
motion for $h_\boldmathsymbol{k}$ reading as
\beq\label{Tensor eq}
h_\boldmathsymbol{k}^{s,\prime\prime} + 
2\mathcal{H}h_\boldmathsymbol{k}^{s,\prime} + k^{2} h^s_\boldmathsymbol{k} = 4 
S^s_\boldmathsymbol{k}, 
\eeq
where $\prime$ denotes differentiation with respect to the conformal time and $s = (+), (\times)$, stands for the two tensor mode polarisation states in general relativity (GR), with $\mathcal{H}$ is the conformal Hubble parameter, while
the polarization tensors  $e^{s}_{ij}(k)$ are the standard  
ones~\cite{Espinosa:2018eve}. The source function 
$S^s_\boldmathsymbol{k}$ is given by~\footnote{We mention here that in this work we neglect possible effects of non-Gaussianities~\cite{Cai:2018dig,Choudhury:2023fwk} and one-loop corrections~\cite{Chen:2022dah} to the induced GW background.}
 \begin{eqnarray}
&&\!\!\!\!\!\!\!\!\!\!\!\!\!\!
S^s_\boldmathsymbol{k}  = \int\frac{\mathrm{d}^3 
\boldmathsymbol{q}}{(2\pi)^{3/2}}e^s_{ij}(\boldmathsymbol{k})q_iq_j\Big[
2\Psi_\boldmathsymbol{q}\Psi_\boldmathsymbol{k-q}  \nonumber\\
&& \!\!\!\!\!\!\!\!
+ 
\frac{4}{3(1\!+\!w_\mathrm{tot})}(\mathcal{H}^{-1}\Psi_\boldmathsymbol{q} 
^{\prime}+\Psi_\boldmathsymbol{q})(\mathcal{H}^{-1}\Psi_\boldmathsymbol{k-q} 
^{\prime}+\Psi_\boldmathsymbol{k-q}) \Big]
\label{eq:Source:def}
\end{eqnarray}
and as one can see is quadratically dependent on the first order scalar perturbation of the metric.
After a lengthy but straightforward calculation, one obtains that the tensor power 
spectrum  $\mathcal{P}^{(s)}_{h}(\eta,k)$ reads as~\cite{Ananda:2006af,Baumann:2007zm,Kohri:2018awv,Espinosa:2018eve}
 \begin{eqnarray}
&&
\!\!\!\!\!\!\!\!\!
\mathcal{P}^{(s)}_h(\eta,k) = 4\int_{0}^{\infty} 
\mathrm{d}v\int_{|1-v|}^{1+v}\mathrm{d}u\! \left[ \frac{4v^2 - 
(1\!+\!v^2\!-\!u^2)^2}{4uv}\right]^{2}\nonumber\\
&&
\ \ \ \ \ \ \ \ \ \ \ \ \ \ \ \ \ \ \ \ \ \ \ \ \ \ \cdot
I^2(u,v,x)\mathcal{P}_\Psi(kv)\mathcal{P}
_\Psi(ku)\,,
  \label{Tensor Power Spectrum}
\end{eqnarray}
where the two auxiliary variables $u$ and $v$ are defined as $u \equiv 
|\boldmathsymbol{k} - \boldmathsymbol{q}|/k$ and $v \equiv q/k$, and the kernel 
function $I(u,v,x)$ is a complicated function containing information for the 
transition between the eMD era driven by the flaton field and the 
lRD era~\cite{Kohri:2018awv,Inomata:2019ivs,Inomata:2019zqy,Papanikolaou:2022chm}. 
Using \Eq{Tensor Power Spectrum}, one can derive the GW spectral abundance, which is defined as follows as~\cite{Maggiore:1999vm}
\beq\label{Omega_GW}
\Omega_\mathrm{GW}(\eta,k)\equiv 
\frac{1}{\bar{\rho}_\mathrm{tot}}\frac{\mathrm{d}\rho_\mathrm{GW}(\eta,k)}{
\mathrm{d}\ln k} = 
\frac{1}{24}\left(\frac{k}{\calH(\eta)}\right)^{2}\overline{\mathcal{P}^{(s)}
_h(\eta,k)},
\eeq
where the bar stands for the oscillation average, i.e the GW envelope.
Finally,  considering that the 
radiation energy 
density reads as $\rho_r = 
\frac{\pi^2}{30}g_{*\mathrm{\rho}}T_\mathrm{r}^4$ and that the temperature of 
the primordial plasma $T_\mathrm{r}$ scales as $T_\mathrm{r}\propto 
g^{-1/3}_{*\mathrm{S}}a^{-1}$, one finds that the GW spectral abundance at 
our present epoch reads as
\beq\label{Omega_GW_RD_0}
\Omega_\mathrm{GW}(\eta_0,k) = 
\Omega^{(0)}_r\frac{g_{*\mathrm{\rho},\mathrm{*}}}{g_{*\mathrm{\rho},0}}
\left(\frac{g_{*\mathrm{S},\mathrm{0}}}{g_{*\mathrm{S},\mathrm{*}}}\right)^{4/3}
\OmegaGW(\eta_\mathrm{*},k),
\eeq
where $g_{*\mathrm{\rho}}$ and $g_{*\mathrm{S}}$ denote the energy and 
entropy relativistic degrees of freedom. Note that the reference 
conformal time $\eta_\mathrm{*}$ in the case of a sudden transition from the eMD to the lRD era should be of 
$\mathcal{O}(1)\eta_r$~\cite{Inomata:2019ivs}.

%%%%%%%%%%%%%%%%%%%%%%%% Section 5: The flipped SU(5) induced gravitational wave signal %%%%%%%%%%%%%%%%%%%%%%%%%%%%%%%%%%%%%%%%%%%%%%%%%%%%%%%%%%%%%%%%

{\bf The flipped SU(5) induced gravitational wave signal}-- 
We are going to focus now on the induced gravitational waves generated within the flipped SU(5) paradigm. Interestingly enough, these induced GWs are significantly enhanced in our case since during the sudden transition from the flaton-driven eMD era to the lRD era, the time derivative of the Bardeen potential goes very quickly from $\Psi^\prime = 0 $ (since in a MD era $\Psi = \mathrm{constant}$ ) to $\Psi^\prime \neq 0$ [See~\cite{Inomata:2019ivs,Domenech:2021ztg} for more details.]. This entails a resonantly enhanced production of GWs sourced mainly by the $\mathcal{H}^{-2}\Psi^{\prime 2}$ term in \Eq{eq:Source:def}. Furthermore, since the sub-horizon energy density perturbations during an MD era scale linearly with the scale factor, i.e. $\delta\propto a$, one should ensure working within the perturbative regime. Thus, we set a non-linear scale $k_\mathrm{NL}$ by requiring that $\delta_{\mathrm{k_\mathrm{NL}}}(\eta_\mathrm{r}) = 1$. In particular, following the analysis of~\cite{Assadullahi:2009nf,Inomata:2020lmk} one can show that the non-linear cut-off scale at which $\delta_{k_\mathrm{NL}}(\eta_r) = 1$ can be recast as
\begin{equation}
k_\mathrm{NL} \simeq \sqrt{\frac{5}{2}}\mathcal{P}^{-1/4}_\mathcal{R}(k_\mathrm{NL})\mathcal{H}(\eta_r).
\end{equation}
Since within the flipped SU(5) paradigm, one predicts a Starobinsky-like inflationary setup with $n_\mathrm{s} = 0.965$~\cite{Ellis:2017jcp}, one can assume as a first approximation a scale-invariant curvature power spectrum of amplitude $2.1\times 10^{-9}$ as imposed by Planck~\cite{Planck:2018vyg}, giving rise to $ k_\mathrm{NL} \simeq 470/\eta_\mathrm{r} \simeq 235 k_\mathrm{r}$~\cite{Inomata:2019ivs},where $k_\mathrm{r}$ is the comoving scale crossing the cosmological horizon at the onset of the lRD era. However, strictly speaking, one should take into account the tilt of the power spectrum leading to a little larger $k_\mathrm{NL}$. Accounting therefore for this tilt effect, we find that $k_\mathrm{NL} \simeq 400k_\mathrm{r}$. 

At this point, we need to mention that one in principle can account for the emission of non-linear modes where $\delta_{k>k_\mathrm{NL}}$, which can potentially enhance the GW signal~\cite{Jedamzik:2010hq}. The treatment of these modes require however high-cost GR numerical simulations, being beyond the scope of the current work. Therefore, by neglecting them we underestimate the GW signal, thus giving a conservative estimate for the GW amplitude.

At the end, the relevant amplified GW signal is calculated as follows 
\beq\label{eq:Omega_GW_resonance}
\Omega^{\mathrm{res}}_\mathrm{GW}(\eta_0,k) = c_g\Omega^{(0)}_\mathrm{r} \Omega_\mathrm{GW}(\eta_\mathrm{lRD},k),
\eeq
where $c_g =\frac{g_{*\mathrm{\rho},\mathrm{*}}}{g_{*\mathrm{\rho},0}}
\left(\frac{g_{*\mathrm{S},\mathrm{0}}}{g_{*\mathrm{S},\mathrm{*}}}\right)^{4/3} \sim 0.4$ and $\eta_\mathrm{lRD}\sim O(\eta_\mathrm{r})$ stands for a time during the lRD era by which the curvature perturbations decouple from the tensor perturbations, thus one  can assume freely propagating GWs while $\Omega_\mathrm{GW}(\eta_\mathrm{lRD},k)$ is derived from \Eq{Omega_GW} where $I(u,v,x)=I_\mathrm{lRD}(u,v,x)$~\cite{Inomata:2019ivs}.

Regarding the frequency of the GW signal, it can be computed as follows
\beq\label{eq:f_GW}
\begin{split}
f_\mathrm{GW} & = \frac{k}{2\pi a_0} = \frac{k}{2\pi a_{\mathrm{d}\Phi}} \frac{a_{\mathrm{d}\Phi}}{a_\mathrm{eq}} \frac{a_\mathrm{eq}}{a_0}  \\ & =
\frac{k}{2\pi a_{\mathrm{d}\Phi}} \left(\frac{\rho_\mathrm{eq}}{\rho_{\mathrm{d}\Phi}}\right)^{1/4}\left(\frac{\rho_\mathrm{0}}{\rho_\mathrm{eq}}\right)^{1/3},
\end{split}
\eeq
where $\rho_\mathrm{eq}=1.096\times 10^{-36}\mathrm{GeV}^4$ and $\rho_0 = 3.6\times 10^{-47}\mathrm{GeV}^4$ are the background energy densities at the matter-radiation equality and today respectively. Lastly, $\rho_{\mathrm{d}\Phi}$ and $a_{\mathrm{d}\Phi}$ respectively stand for the energy density and the scale factor at the end of the eMD, namely at the time of the flaton decay.

In order to compute now $\rho_{\mathrm{d}\Phi}$ and $a_{\mathrm{d}\Phi}$, let us recap the basic physical quantities describing the dynamics of the flaton-driven eMD era. In particular, the Hubble parameter during the flaton-dominated era is given by~\cite{Ellis:2018moe} 
\beq
H \;=\; \left(\frac{\rho_{\Phi}}{3M_P^2}\right)^{1/2} \;=\; \left(\frac{\zeta(3)m_{\Phi}T_{\Phi}^3}{3\pi^2 M_P^2}\right)^{1/2}\,.
\eeq

With regard now to the decay rate $\Gamma_\Phi$ of the flaton, the latter was calculated in~\cite{CAMPBELL1987355} via effective $D$-term diagrams, leading to 
\beq
\Gamma_{\Phi}\;\simeq\; \frac{9\lambda_{1,2,3,7}^4}{2048\pi^5}\left(\frac{m_{\Phi}m_{F,\bar{f},\ell^c,\tilde{\phi}_a}^2}{M_{\rm GUT}^2}\right)\,
\eeq
where $m_\Phi$ is the mass of the flaton, $M_\mathrm{GUT}$ is the GUT energy scale and $m_{F,\bar{f},\ell^c,\tilde{\phi}_a}$ are the masses of the various fields and $\lambda_{1,2,3,7}$ are appropriate Yukawa couplings of the order of $O(1/2)$ [See \cite{Ellis:2018moe} for more details.]. Assuming to a very good approximation $m_{F,\bar{f},\ell^c,\tilde{\phi}_a}\simeq m_\Phi$~\cite{Ellis:2018moe}, one finds that the flaton decays approximately when $H\sim\Gamma_{\Phi}$, or equivalently when the flaton temperature is
\beq\label{eq:T_dPhi}
T_{d\Phi} \;\simeq\; \frac{3\lambda_{1,2,3,7}^{8/3}}{128}\left(\frac{9}{2\zeta(3)\pi^8}\right)^{1/3}\left(\frac{m_{\Phi}m_{F,\bar{f},\ell^c,\tilde{\phi}_a}^4 M_P^2}{M_{\rm GUT}^4}\right)^{1/3}\,.
\eeq

At this point, we should stress that the flaton field decays through effective D-term diagrams into supersymmetric particles at a temperature $T_{\mathrm{d}\Phi}$ of around 1-10$\mathrm{keV}$ [See \Eq{eq:T_dPhi}], leading at the end to a peak frequency of the order of $\mathrm{nHz}$ as shown in \Eq{Frequency_Phen} and \Eq{eq:f_GW_peak}. Then, the flaton decay products decay quickly into the truly relativistic Standard Model particles which thermalize at a temperature $T_\mathrm{reh}\sim\mathcal{O}(\mathrm{MeV})$, being higher than  $T_{\mathrm{d}\Phi}$, producing in this way an amount of entropy necessary to account for the present-day baryon asymmetry. One then recovers the standard model plasma at a temperature of a few $\mathrm{MeV}$, around what is needed to start the process of nucleosynthesis. See~\cite{Ellis:2018moe,Ellis:2019opr} for more details.

Having reviewed above the dynamics of the flaton-dominated era and recap its decay process, let us derive here the characteristic scales of the problem at hand. In particular, the mode $k_\Phi$, which crosses the cosmological horizon at the onset of the flaton domination era at $T\sim m_\Phi$, will be written as $k_\Phi = a_\Phi H_\Phi = \left(\frac{\zeta(3)m^4_\Phi}{3\pi^2\Mp^2}\right)^{1/2}$, where we have normalised $a_\Phi=1$ whereas the mode $k_{\mathrm{d}\Phi}$ crossing the horizon at the time of the flaton decay when $H\sim \Gamma_\Phi$ will read as
$k_{\mathrm{d}\Phi} = a_{\mathrm{d}\Phi}H_{\mathrm{d}\Phi}$, where
$H_{\mathrm{d}\Phi}= \Gamma_\Phi$ and $a_{\mathrm{d}\Phi} =  (H^ 2_\Phi/H^ 2_{\mathrm{d}\Phi})^{1/3}$.

Finally, the peak frequency of our problem at hand, where one expects a resonantly amplified GW signal, is associated with the non-linear comoving cut-off scale, $k_\mathrm{NL}=400k_{r}$, where in our case $k_\mathrm{r}=k_{\mathrm{d}\Phi}$. At the end, plugging $k_{\mathrm{d}\Phi}$ in \Eq{eq:f_GW}, we obtain that the peak frequency $f_\mathrm{GW,peak}$ reads as
\beq
\begin{split}
& f_\mathrm{GW,peak}  = \frac{k_\mathrm{NL}}{2\pi a_0} = \frac{k}{2\pi a_{\mathrm{d}\Phi}} \frac{a_{\mathrm{d}\Phi}}{a_\mathrm{eq}} \frac{a_\mathrm{eq}}{a_0}  \\ & =
\frac{k}{2\pi a_{\mathrm{d}\Phi}} \left(\frac{\rho_\mathrm{eq}}{\rho_{\mathrm{d}\Phi}}\right)^{1/4}\left(\frac{\rho_\mathrm{0}}{\rho_\mathrm{eq}}\right)^{1/3} \\ & = 1.5\times 10^{-9}\left(\frac{\lambda_{1,3,5,7}}{0.5}\right)^2\left(\frac{m_\Phi}{10^4\mathrm{GeV}}\right)^{3/2}\left(\frac{10^{16}\mathrm{GeV}}{M_\mathrm{GUT}}\right). \label{Frequency_Phen}
\end{split}
\eeq

\begin{figure}[h!]
\begin{center}
\includegraphics[width=0.52\textwidth]{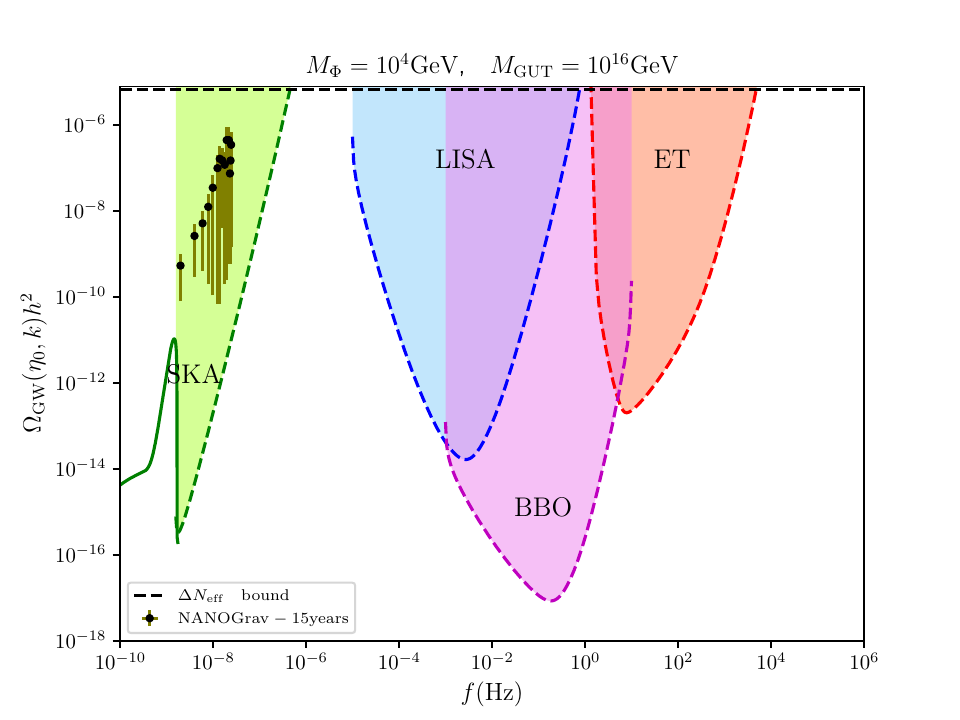}
\caption{ 
{\it{The primordial GW background non-linearly induced by inflationary perturbations (solid green curve) arising from flipped $\mathrm{SU(5)}$ superstring theory for $m_\mathrm{\Phi} = 10^4 \mathrm{GeV}$ and $M_{\mathrm{GUT}} = 10^{16} \mathrm{GeV}$ }. On the top of our theoretical prediction for the induced stochastic GW background we show the  15-year NANOGrav GW data, as well as the sensitivities of SKA~\cite{Janssen:2014dka}, LISA~\cite{LISACosmologyWorkingGroup:2022jok,Karnesis:2022vdp}, BBO~\cite{Harry:2006fi} and ET~\cite{Maggiore:2019uih} GW experiments. In the horizontal black dashed line, we show also the upper bound on $\Omega_\mathrm{GW,0}h^2\leq 6.9\times 10^{-6}$ coming from the upper bound constraint on $\Delta N_\mathrm{eff}$ from CMB and BBN observations~\cite{Smith:2006nka}.} 
}
\label{fig:GW_signals}
\end{center}
\end{figure}

Remarkably, within the framework of flipped $\mathrm{SU(5)}$ one can re-express the mass parameters of \Eq{Frequency_Phen} in terms of solely the Planck Mass and ultimately directly in terms of only the string slope parameter $\alpha'$. In particular, as it was shown in \cite{Ellis:2018moe, Ellis:2019opr}, there is a relation between the flaton vev, $\langle \Phi \rangle $, and the flaton soft supersymmetry-breaking mass term $m_\Phi$
\beq
|m_\Phi | \sim \frac{\langle \Phi \rangle^6}{M^5_\mathrm{Pl}}.
\eeq
Since the flaton vev is of order of the $\mathrm{GUT}$ scale $\langle \Phi \rangle \approx M_{\mathrm{GUT}}$~\cite{Ellis:2018moe}, then, by substituting $M_\mathrm{GUT}= 10^{16} \mathrm{GeV}$ and $\Mp = 4M_\mathrm{s}$  where $M_\mathrm{s}$ is the heterotic string scale being defined as $M_\mathrm{s}\equiv 1/\sqrt{2 \alpha'}$ \cite{Antoniadis:2020txn} one obtains that 
\beq\label{eq:f_GW_peak}
f_\mathrm{GW,peak}= 1.5 \times 10^{-9} \left(\frac{\lambda_{1,2,3,7}}{0.5}\right)^2\left(\frac{\alpha'}{8/\Mp^2}\right)^4 \mathrm{Hz}.
\eeq 

We note here that in our analysis we ignored the GWs which are resonantly enhanced during the first reheating stage when the inflaton field oscillates at the bottom of its potential and during which the Universe effectively behaves as pressureless matter~\cite{Ellis:2017jcp,Ellis:2018moe}. During this first eMD era before the flaton field dominates the Universe content, GWs can be abundantly produced at higher frequencies compared to the flaton-driven eMD era induced GWs. However, since after the end of the first reheating stage there follows a long second eMD era driven by the flaton, one expects that these GWs are sufficiently diluted since $\Omega_\mathrm{GW}\propto \rho_\mathrm{GW}/\rho_\mathrm{flaton}\propto a^{-4}/a^{-3}\propto 1/a$. The relevant dilution factor of these primordial GWs generated during the first reheating stage will read as 
$\Delta \equiv \frac{a_{\mathrm{d}\Phi}}{a_\Phi} \simeq 10^{10}$,
for $m_\Phi = 10^{4}\mathrm{GeV}$ and $M_\mathrm{GUT}=10^{16}\mathrm{GeV}$. Nevertheless, if one accounts for the non-linear growth of the perturbations entering the inflaton-driven eMD era, they may enhance the GW amplitude rendering it potentially detectable by the LISA, BBO and ET GW probes.

%Finally, we should stress that the flaton field decays completely into the SM particles through effective D-term diagrams at a temperature of a few $\mathrm{MeV}$~\cite{CAMPBELL1987355,Ellis:2017jcp,Ellis:2018moe,Ellis:2019opr}, hence not contributing to the count of dark relativistic species constrained by $\Delta N_\mathrm{eff}$ bounds at BBN. Thus, the early matter era driven by the flaton field, which is present here, is compatible with the standard picture of the primordial BBN.

%%%%%%%%%%%%%%%%%%%% Section 6: Conclusions %%%%%%%%%%%%%%%%%%%%%%%%%%%%%%%%%%%%%%
{\bf Conclusions} -- 
The flipped SU(5) superstring paradigm accommodates a very rich particle and cosmological phenomenology consistent with observations. In particular, it can naturally give rise to stable Starobinsky-type inflation preferred by Planck. Whats more, it can explain successfully the lepton and quark mass, while at the same time it can provide a realistic mechanism for the generation of the baryon asymmetry in the Universe and the production of neutralino dark matter. Among its most interesting aspects is its strong reheating scenario which proceeds via the incoherent component of the flaton field.

In this Letter, we focused on a particular phenomenology associated with the flaton field which naturally arises in the flipped SU(5) superstring paradigm. Specifically, as it was shown in~\cite{Ellis:2017jcp,Ellis:2018moe}, one meets an eMD era driven by the flaton which transitions quite suddenly to the lRD era. This sudden transition leads to an abundant production of a GW signal non-linearly induced by inflationary curvature perturbations, constituting a pure observational GW signature of the flipped SU(5) superstring theory. Remarkably, we found that the peak frequency of this GW signal $f_\mathrm{GW,peak}$ depends only on the string slope $\alpha'$ and reads as 
\beq\label{eq:f_GW_peak_conclusions}
f_\mathrm{GW,peak} \propto 10^{-9} \left(\frac{\alpha'}{\alpha'_*}\right)^4 \mathrm{Hz},
\eeq
where $\alpha'_*$ is the fiducial string slope being related to the reduced Planck scale $\Mp$ as $\alpha'_* = 8/\Mp^2$. Notably, as one may see from \Eq{eq:f_GW_peak_conclusions}, the peak frequency of these GWs lies in the nHz frequency range and is very close to the NANOGrav/PTA data; hence rendering this primordial GW signal potentially detectable by SKA, NANOGrav and PTA probes at their very low frequency region of their detection bands. 

If one now takes into account via GR numerical simulations the highly non-linear energy density perturbations present at smaller scales, i.e. at higher frequencies~\cite{Jedamzik:2010hq}, the expected GW signal will be significantly enhanced at these scales compared to the one given here and displaced within the $10^{-9}-10^{-8}\mathrm{nHz}$ frequency range, hence being potentially a good candidate to explain the NANOGrav/PTA GW signal [See here for recent works~\cite{Bringmann:2023opz,Fujikura:2023lkn,Niu:2023bsr,HosseiniMansoori:2023mqh,Choudhury:2023kam,Wu:2023pbt,Das:2023nmm,Basilakos:2023xof,Yi:2023npi,Datta:2023xpr,Balaji:2023ehk,Gouttenoire:2023nzr,Bhaumik:2023wmw,Franciolini:2023wjm,Franciolini:2023pbf,Inomata:2023drn,Inomata:2023zup,Datta:2023vbs,Bernardo:2023zna,Choudhury:2023fwk,Huang:2023chx} on different cosmological models explaining the NANOGrav/PTA signal.]. Lastly, due to the linear growth of the sub-horizon energy density perturbations during the second reheating stage driven by the flaton, one expects the formation of flaton structures similar to the inflaton ones~\cite{Jedamzik:2010dq,Hidalgo:2022yed} as well as primordial black holes~\cite{Martin:2019nuw,Martin:2020fgl} with very interesting phenomenology. These aspects will be studied in future works. 

{\bf Acknowledgements} -- 
The work of DVN was supported
in part by the DOE grant DE-FG02-13ER42020 at Texas A\&M University and in part by the Alexander S. Onassis Public Benefit Foundation.
SB, ENS, TP and CT acknowledge the 
contribution of the LISA CosWG and the COST Actions  CA18108 ``Quantum Gravity Phenomenology in the multi-messenger approach''  and 
CA21136 ``Addressing observational tensions in cosmology with systematics and 
fundamental physics (CosmoVerse)''. TP and CT acknowledge as well financial support from the Foundation for Education and European Culture in Greece and A.G. Leventis Foundation respectively.

\bibliography{ref}

\end{document}